\documentclass[11pt,twoside]{article}
\usepackage{asp2010}

\begin{document}

\title{An empirical clock to measure the dynamical age of stellar systems.}
\author{Emanuele Dalessandro$^1$
\affil{$^1$ Dipartimento di Fisica e Astronomia, Universit\'a degli Studi di Bologna, Viale
Berti Pichat 6/2, I-40127 Bologna, Italy}}

\begin{abstract}
Blue Straggler Stars (BSS) are among the brightest and more massive stars in globular clusters (GCs).
For this reason they represent an ideal tool to probe the dynamical evolution of these stellar systems.
Here I show, following the results by Ferraro et al. (2012), that the BSS radial distribution 
can be used as a powerful indicator of the cluster dynamical age. In fact on the basis of their BSS radial distribution shape, 
GCs can be efficiently grouped in different families corresponding to the different dynamical stages reached by 
the stellar systems. This allows to define a first empirical clock, {\it the dynamical clock}, able 
to measure the dynamical age of a stellar system from pure observational quantities.

\end{abstract}

\section{Introduction}

Among the large variety of exotic objects which populate the dense environments of globular clusters (GCs), 
for sure Blue Straggler Stars (BSS)
represent the most numerous and ubiquitous population. BSS were observed for the first time by Sandage (1953)
in the outer regions of M3. Since then, they have been detected in any properly observed stellar system,
from open clusters (see for example Mathieu \& Geller 2009) to dwarf galaxies (Mapelli et al. 2009).\\
In the color magnitude diagram (CMD) of an old stellar population, BSS define a sparsely populated sequence more luminous and bluer than the turn-off
(TO) mass of a normal hydrogen-burning main sequence (MS) star. Therefore they appear younger and more
massive than normal cluster stars. Indeed observations have shown that they have a mass $m=1.2-1.7
M_{\odot}$ (Shara et al. 1997; Gilliland et al. 1998; De Marco et al. 2004), which is about twice that 
of TO stars ($m\sim0.8M_{\odot}$), thus they are thought to be the result of some mechanism responsible for increasing the
masses of single stars. \\
Two main formation scenarios have been proposed over the years: the mass transfer scenario 
(MT-BSS; McCrea 1964; Zinn \&
Searle 1976) according to which BSS are the result of mass accretion between two stars in a
binary system, and the collisional scenario (COL-BSS; Hills \& Day 1976) in which BSS are the end products
of stellar mergers induced by collisions between single stars or binary systems.
These mechanisms are believed to work simultaneously within the same cluster (see the case of M30 - Ferraro et al. 2009 - 
and NGC362 - Dalessandro et al. 2013a), with efficiencies that may be function of the
environment (Ferraro et al. 1995; Davies et al. 2004).\\
Independently of the formation mechanism and because of their mass, BSS are heavily affected by dynamical friction 
and thus they are natural test particles to probe the internal dynamics of stellar aggregates.
In particular their radial distribution has been found to be an important tool to probe the efficiency of dynamical
friction and the dynamical age of stellar systems (Ferraro et al. 2012, hereafter F12).

\begin{figure}
\plottwo{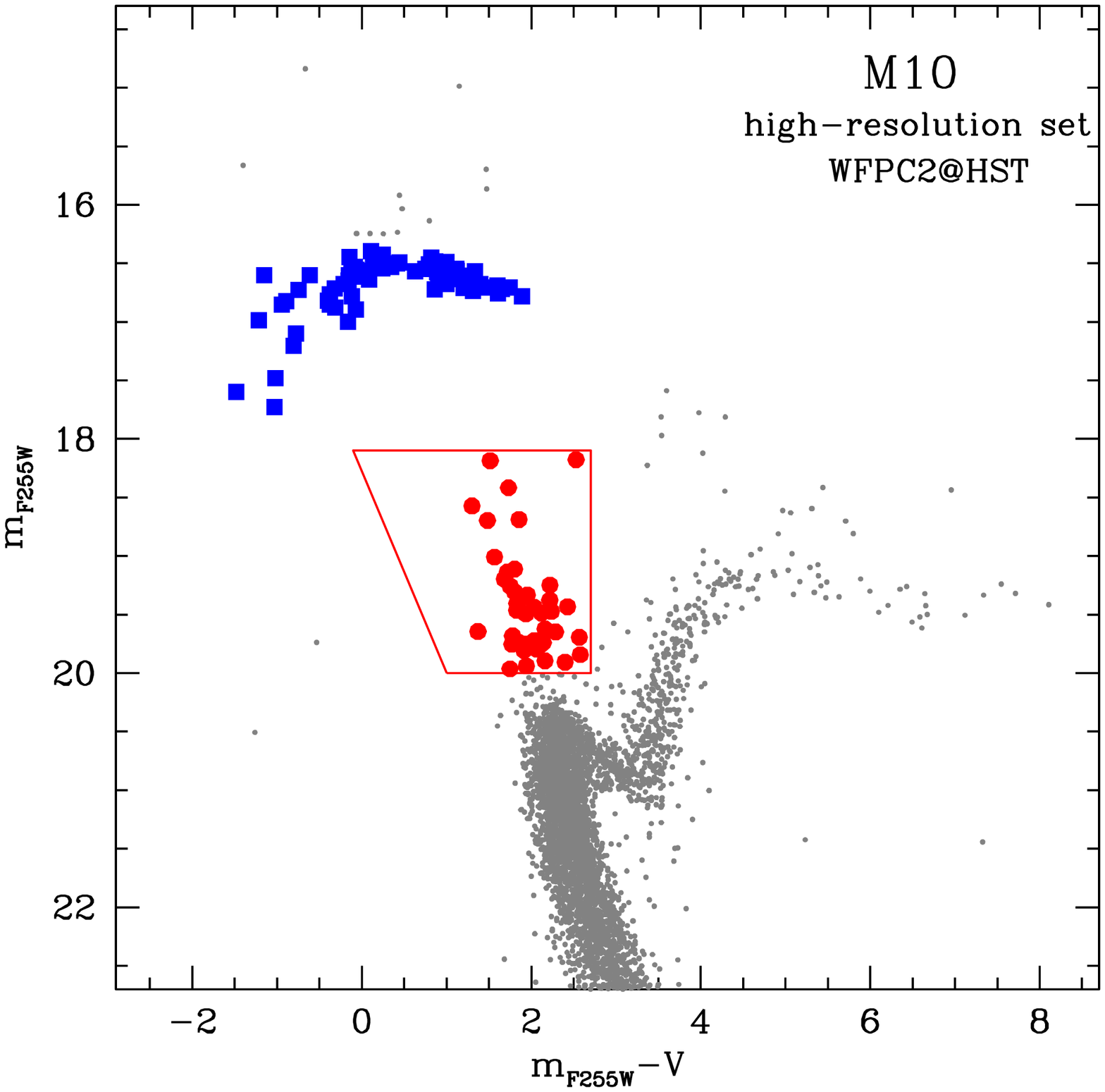}{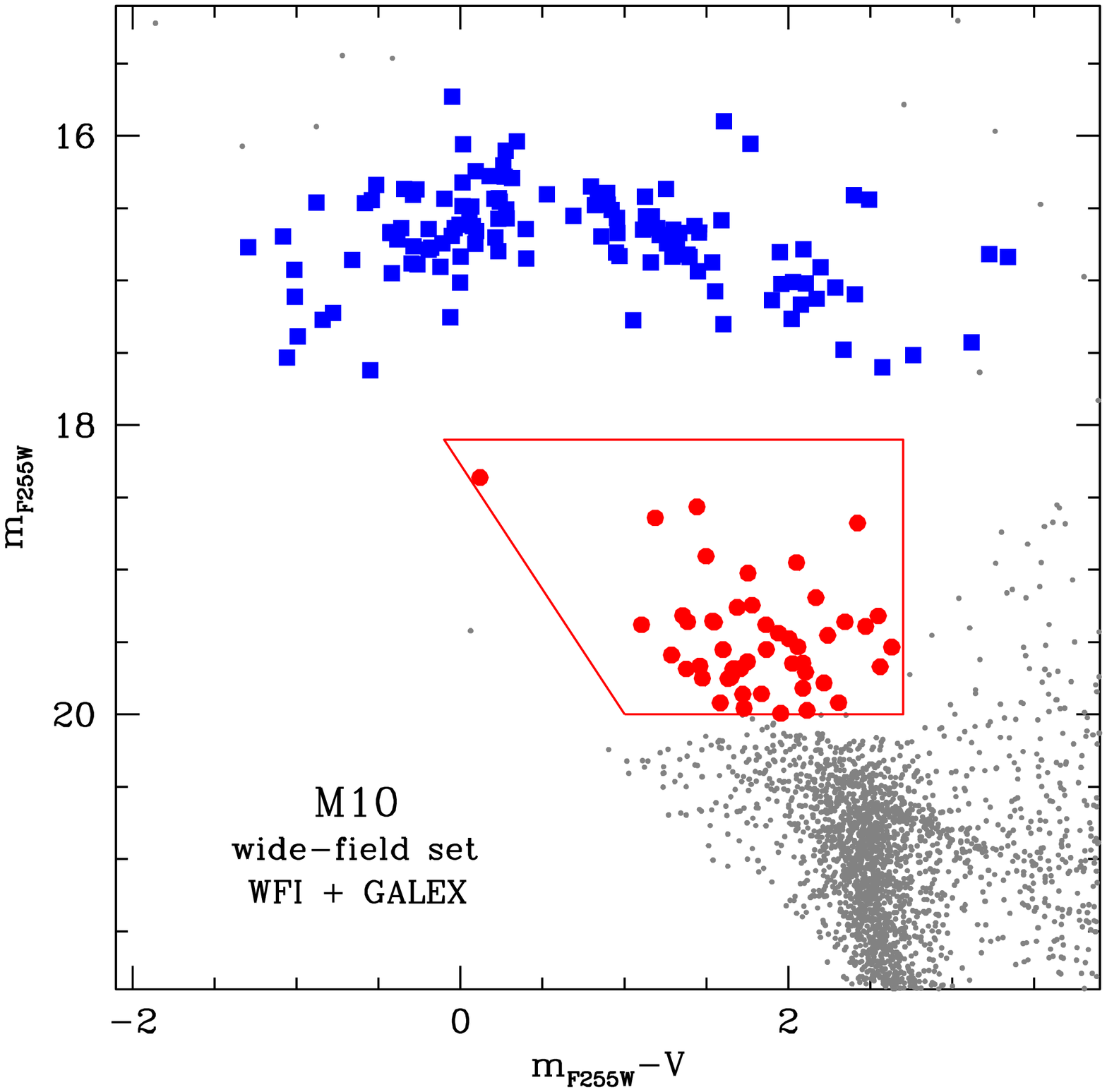}
\caption{UV CMDs of the GC M10 (data from Dalessandro et al. 2013b). The left panel shows 
the high-resolution set obtained with the Wide Field Planetary Camera 2 (WFPC2) aboard HST. 
In the right panel the
wide-field set, which is the result of a proper combination of optical data obtained with 
the Wide Field Imager (WFI) mounted at the ESO-MPG telescope and GALEX. 
The selected BSS are highlighted with red circles, while the HB
with blue squares.  
}
\label{}
\end{figure}

\section{The general approach}

In order to obtain a reliable BSS radial distribution it is important to build a complete sample of BSS and 
of at least one reference population (typically Red Giant Branch - RGB - or Horizontal Branch - HB - stars)
for the entire
cluster extension (see for example Ferraro et al. 1993; 1997).
With this aim, our group have observed, over the past fifteen years, about 25 GCs 
adopting a multi-band and multi-telescope approach (see Dalessandro et al. 2008 and references therein).
In particular, we combined (Figure~1) high-resolution optical and ultra-violet (UV) Hubble Space Telescope (HST) data 
to properly sample the innermost
and most crowded regions, and wide-field optical (from ground-based facilities like ESO, CFHT, SUBARU and LBT; see
Beccari et al. 2013 for example) 
and UV (from the space mission GALEX; Schiavon et al. 2012, Dalessandro et al. 2012) data to observe the external
regions and sample the entire extension of the clusters. Typical examples of the adopted procedures are described
by Dalessandro et al. (2009; 2013b) for the cases of M2 and M10. \\
It is important to stress that the use of UV filters is of crucial importance for systematic studies of BSS.
In fact, the optical CMD of old stellar populations is dominated by the cool stellar component, as a consequence 
the construction of complete samples of hot stars (like BSS or extreme HB stars) is difficult in this plane. 
Moreover in the optical plane, BSS may be contaminated by blends of Sub Giant Branch (SGB) and RGB stars. 
On the contrary at UV wavelengths, BSS are among the brightest sources (see Figure~1), while RGB are faint. 
In particular, as shown in Figure~1, BSS describe an almost narrow sequence spanning 2-3 mag in the UV CMD, and so
they are more easily detectable. In addition, blends are less severe because of the relative faintness of RGB and 
SGB. Indeed CMDs involving UV bands, like the HST F255W filter, are ideal tools for selecting BSS even in the
cores of the densest GCs. 

\section{The BSS radial distribution}

Once complete samples are built, it is possible to study the BSS radial distribution by dividing the surveyed clusters
in a number of concentric annuli and counting in each of them the number of BSS and reference populations (RGB and HB)
stars.
In this way it is possible to use the double normalized ratio $R_{POP}$ defined by Ferraro et al. (1993).
The radial distribution of HB and RGB stars follows that of the cluster sampled light and $R_{HB/RGB}=1$ as expected by
stellar evolution theory (Renzini \& Buzzoni 1986) for post-MS stars. 
On the contrary the BSS radial distribution shows different and more complex behaviors.

\begin{figure}
\centering
\includegraphics[scale=0.4]{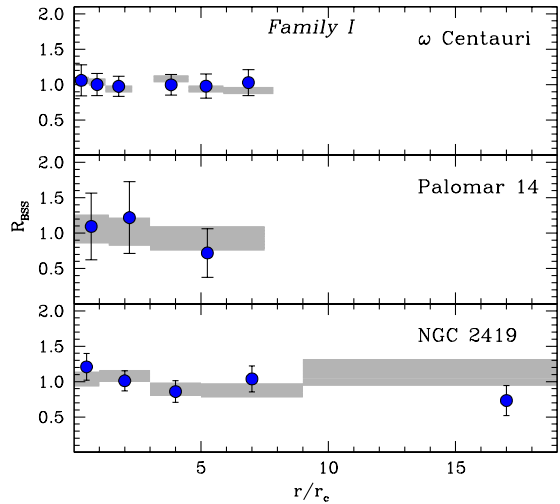}
\caption{The radial distribution of BSS in dynamical young ({\it Family I}) GCs (F12). Blue circles correspond to
the BSS double normalized ratio 
ratio ($R_{BSS}$) as a function of the distance from the cluster center normalized to the 
cluster core radius. The grey shaded areas correspond to the double normalized ratio for the reference 
population (RGB or HB).}
\label{map}
\end{figure}

\begin{figure}
\centering
\includegraphics[scale=0.4]{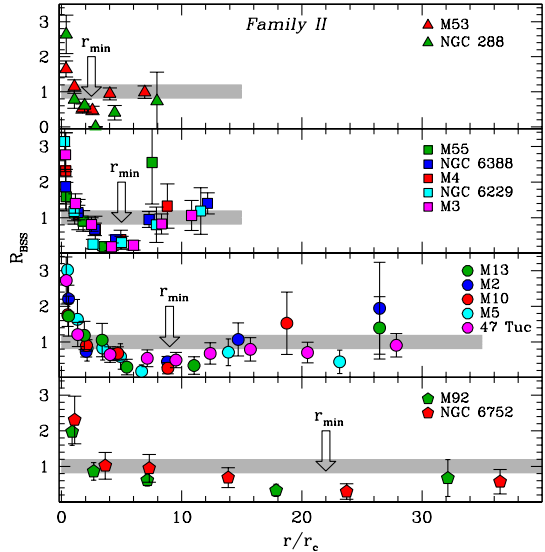}
\caption{As in Figure~2, but for intermediate dynamical-age ({\it Family II}) GCs.}
\label{map}
\end{figure}

F12 have analyzed the entire database of BSS collected by our group.
This data-base is made of clusters with almost the same age ($t\sim12-13$ Gyr)\footnote{With the only exception 
of Palomar~14 (which has $t=10.5$ Gyr)} and very different structural properties. Although significant variations in the
radial distribution of BSS between clusters are already known, F12 found that, when the radial distance is 
expressed in units
of the core radius ($r_c$), the BSS radial distribution is
surprisingly similar within distinct subsamples. Similarities are so evident that all the clusters can be grouped in at
least three different families. The BSS radial distributions ($R_{BSS}$ compared to $R_{HB/RGB}$) of the three groups 
of clusters defined by F12 are shown in Figures 2, 3 and 4. As can be seen from these plots, most of the clusters 
show a bimodal BSS radial distribution: highly peaked in the center, with a clear-cut dip at
intermediate radii and with an upturn in the external regions (Figure~3). However some exceptions to this general
behavior exist. In fact, in
some clusters, BSS have the same radial distribution of the reference populations (Figure~2), while in others BSS
show a monotonic decreasing distribution (Figure~4).

\section{The dynamical clock}

Simple analytic models (Mapelli et al. 2004, 2006) have shown that the BSS radial distribution is primarily modeled 
by the long-term effect of dynamical friction acting on the cluster binary population and its by-products.  
In fact, whereas BSS generated by stellar collisions are expected to be the main contributors to the central peak of the
distribution (Davies et al. 2004), the portion beyond the cluster core, where the minimum is observed (Figure~3) is due to
BSS generated by mass-transfer or merger in primordial binary systems (Geller \& Mathieu 2011). MT-BSS are the by-product
of the evolution of a $\sim1.2 M_{\odot}$ primordial binary that has been orbiting the cluster and suffering the effects of
dynamical friction acting at larger and larger distances from the cluster center. Therefore binary systems are 
expected to drift
toward the cluster core and their radial distribution is expected to develop a central peak in the cluster
center and a dip that progressively propagates outwards. As the dynamical evolution of the host systems proceeds, the
portion of the cluster where dynamical friction has been effective increases and the position of the minimum in the radial
distribution ($r_{min}$) increases. Interestingly, there is a good agreement between the location
of $r_{min}$ and the theoretical expected value (Binney \& Tremaine 1987) of the radius at which the dynamical 
friction is expected to segregate the BSS (Figure~3).\\ 
Following this interpretation, the groups of clusters shown in Figures 2, 3 and 4 corresponds to families with different
dynamical ages. In particular, clusters with a flat radial distribution ({\it Family I}, Figure~2) are dynamically young,
so that the effect of dynamical friction is still not present. In more evolved clusters ({\it Family II}; Figure~3),
dynamical friction starts to be effective and segregates heavy stars orbiting at distances relatively close to the center. 
As a consequence a peak in the center and a minimum at small radii appear in the BSS distribution. Meanwhile, 
the most remote BSS have not yet been affected by the action of dynamical friction, thus generating the external rising
branch and the bimodal distribution. With time, dynamical friction extends its action at progressively larger distances
from the cluster center, as a consequence the dip in the BSS radial distribution moves outwards (note the different positions of
$r_{min}$ in the four panels of Figure~3). Indeed in highly
dynamically evolved clusters ({\it Family III}; Figure~4) also the most external BSS are expected to have already sunk
toward the cluster center and the rising branch to have already disappeared.

\begin{figure}
\centering
\includegraphics[scale=0.4]{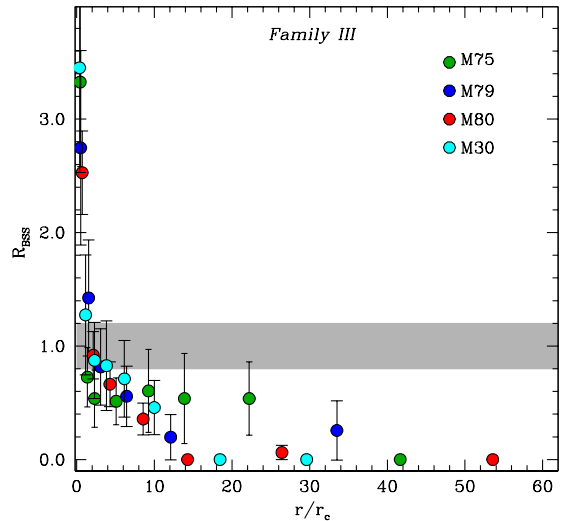}
\caption{As in Figure~2, but for dynamically old systems ({\it Family III}).}
\label{map}
\end{figure}

The classification in three groups of GCs based on their dynamical age is of course a first approximation. 
Indeed $r_{min}$ is
expected to move continuously. On this basis, F12 defined the first empirical clock, {\it the dynamical clock}, able 
to measure the dynamical age of a stellar system from pure observational quantities. $r_{min}$ is the clock hand 
of the {\it dynamical clock}. 
In fact as dynamical friction moves $r_{min}$ within the cluster, the location of $r_{min}$ allows the observer
to measure the dynamical age of the system.\\
As expected for a meaningful {\it dynamical clock}, $r_{min}$ nicely anti-correlates with the central relaxation time
($t_{rc}$): clusters with relaxation times of the order of the age of the Universe ($t_H=13.7$ Gyr) 
show no signs of BSS segregation, hence the radial distribution of BSS is flat and $r_{min}$ is not definable. For
decreasing relaxation times values, $r_{min}$ increases progressively.
$t_{rc}$ is indicative of the relaxation time at a specific radial distance from the cluster center. On the
contrary the {\it dynamical clock} defined by F12 is much more sensitive to the global dynamical evolutionary stage
reached by the cluster. In fact the BSS radial distribution is able to simultaneously probe the dynamical evolution
degree at all distances from the cluster center, providing a much finer ranking of dynamical ages.

\begin{figure}
\centering
\includegraphics[scale=0.4]{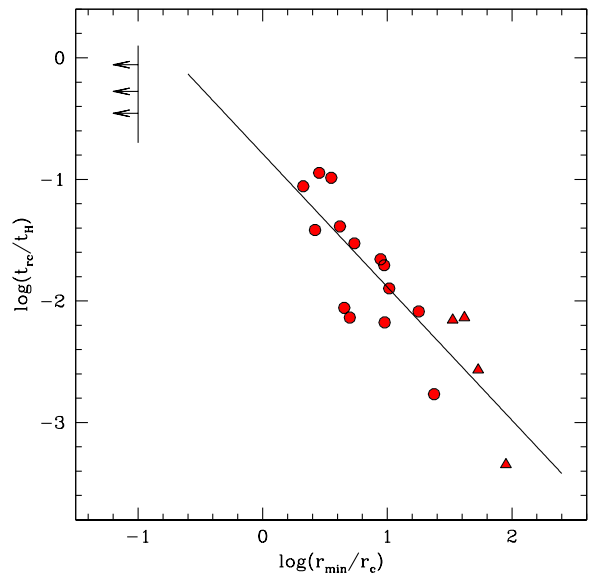}
\caption{Central relaxation time ($t_{rc}$), normalized to the age of the Universe $t_H$,
as a function of $r_{min}$, expressed in units of the core radius $r_c$. 
{\it Family II} systems are plotted as filled circles.
{\it Family I} clusters are plotted as lower-limit arrows at $r_{min}=0.1$. 
Triangles correspond to the dynamically old ({\it Family III}) GCs. 
A clear anti-correlation is found: GCs with relaxation times of the order of $t_H$ show no signs of
BSS segregation, while for decreasing relaxation times the radial position of the minimum increases progressively.}
\label{map}
\end{figure}

\acknowledgements This research is part of the project {\it COSMIC-LAB} 
(http://www.comic-lab.eu) funded by the {\it European Research Council} (under contract ERC-2010-AdG-267675).


\end{document}